# Dynamic Resource Allocation for Virtual Machine Migration Optimization using Machine Learning

**Yulu Gong[1]\*, Jiaxin Huang[2], Bo Liu[3], Jingyu Xu[4], Binbin Wu[5], Yifan Zhang[6]**

1 Computer & Information Technology, Northern Arizona University, Flagstaff, AZ, USA
2 Information Studies, Trine University, Phoenix USA
3 Software Engineering, Zhejiang University, HangZhou China
4 Computer Information Technology, Independent Researcher, Flagstaff, AZ, USA
5 Heating Ventilation and Air Conditioning Engineering, Tsinghua University, Beijing China
6 Computer & Information Technology, Northern Arizona University, Flagstaff, AZ, USA
\* Corresponding author: Bo Liu [E-mail:yg486@nau.edu]

## ABSTRACT

The paragraph is grammatically correct and logically coherent. It discusses the importance of mobile terminal cloud computing migration technology in meeting the demands of evolving computer and cloud computing technologies. It emphasizes the need for efficient data access and storage, as well as the utilization of cloud computing migration technology to prevent additional time delays. The paragraph also highlights the contributions of cloud computing migration technology to expanding cloud computing services. Additionally, it acknowledges the role of virtualization as a fundamental capability of cloud computing while emphasizing that cloud computing and virtualization are not inherently interconnected. Finally, it introduces machine learning-based virtual machine migration optimization and dynamic resource allocation as a critical research direction in cloud computing, citing the limitations of static rules or manual settings in traditional cloud computing environments. Overall, the paragraph effectively communicates the importance of machine learning technology in addressing resource allocation and virtual machine migration challenges in cloud computing.

## 1. INTRODUCTION

In today's digital age, cloud computing technology has become a key tool for enterprises and organizations to achieve flexibility, efficiency and scalability. With the widespread application of cloud computing, the deployment and management of virtual machines in cloud environments become more and more complex. In this complex environment, efficient resource allocation and virtual machine migration optimization have become important challenges in the cloud computing field. Traditional methods of resource allocation and virtual machine migration often rely on static rules or manual Settings, which are difficult to adapt to the dynamic changes of the cloud environment, resulting in resource waste and performance degradation. Therefore, how to use advanced technical means, such as machine learning, to achieve dynamic resource allocation and virtual machine migration optimization has become an urgent problem in the field of cloud computing.

In the cloud computing environment, virtual machine migration is a key technology, and its importance is self-evident. First, virtual machine migration enables dynamic resource adjustment and load balancing, thereby improving utilization and performance of the entire cloud environment. Secondly, VM migration can maintain, expand, or recover from faults in the cloud environment without affecting user services, improving the availability and reliability of the system. However, VM migration also faces many challenges, including performance loss during migration, length of migration time, and possible data consistency issues during migration.

At present, the traditional optimization of resource allocation and virtual machine migration often relies on static rules or manual Settings, which has some problems and limitations. First, static rules often fail to adapt to the dynamic changes in the cloud environment, resulting in insufficient or excessive resource allocation, which affects the performance and utilization of the system. Secondly, because the VM migration process needs to consider a variety of factors, such as network bandwidth and host load, static rules are often unable to flexibly cope with, resulting in low migration efficiency and system performance fluctuations. In addition, the potential for data loss or data consistency issues during VM migration is one of the limitations of the current approach.

In view of the above problems and limitations, the use of machine learning technology for dynamic resource allocation and virtual machine migration optimization has become a potential solution. Machine learning technology can automatically discover the pattern and trend of resource utilization through the learning and analysis of a large number of historical data, so as to realize the dynamic adjustment and optimization of resources. In terms of virtual machine migration, machine learning technology can use real-time monitoring data and predictive models to intelligently select the best migration solution, reducing performance losses and data consistency issues during migration. Therefore, dynamic resource allocation and virtual machine migration optimization using machine learning technology can better adapt to the dynamic changes of the cloud environment and improve the performance and reliability of the system.

## 2. BACKGROUND AND RELATED WORK

### 2.1 VM Migration in the Cloud Computing Environment

In the rapidly evolving landscape of cloud computing, virtual machine (VM) migration plays a pivotal role in ensuring the agility, scalability, and efficiency of cloud infrastructures. Virtual machine migration, facilitated by virtualization technology, entails the seamless transfer of running VM instances from one physical server to another within the cloud environment. This technique optimizes resource utilization, enhances fault tolerance, and facilitates load balancing across servers.

The process of VM migration typically involves several key steps to ensure the continuity of service and minimal disruption to operations. Firstly, the source server initiates the migration by freezing the state of the VM, including its memory, disk, and network connections. Subsequently, the source server transfers the frozen state, memory, and disk data to the target server. Upon receiving the data, the target server instantiates a new VM instance and loads the transferred memory and disk data into it. To maintain uninterrupted network connectivity, the network connection is switched from the source server to the target server. Finally, the state of the VM is unfrozen on the target server, allowing it to resume normal operation seamlessly.

Despite its benefits, VM migration poses several challenges that must be addressed for successful execution. One of the primary challenges is ensuring minimal disruption to ongoing operations during the migration process. Factors such as network bandwidth, latency, and resource availability between the source and destination servers must be carefully considered to optimize performance and mitigate potential downtime. Additionally, choosing the appropriate migration timing is crucial to avoid impacting critical applications and services running on the VMs.

In conclusion, VM migration in the cloud computing environment is a fundamental technique for optimizing resource utilization, enhancing fault tolerance, and ensuring scalability. By understanding and addressing the challenges associated with VM migration, cloud providers can effectively leverage this technique to deliver seamless and efficient services to their users.

### 2.2 Resource allocation in the cloud computing environment

Rational allocation of resources plays a crucial role in cloud computing. In the resource allocation of cloud computing, the cloud computing center has limited cloud resources, and users arrive in sequence. Each user requests the cloud computing center to use a certain number of cloud resources at a specific time. The cloud computing service center needs to decide whether to accept the user's request or put the user's request on hold. Reasonable allocation of cloud resources can maximize the quality of service users and improve user satisfaction under limited resources. Most of the existing resource allocation algorithms use heuristic algorithms to allocate resources, but it is difficult for heuristic algorithms to achieve good results in complex situations.

The resource allocation of cloud computing needs to consider various needs of users. Different users have different requirements on cloud resources. The cloud computing center needs to allocate resources according to user requirements.

Users' demand for cloud resources can be divided into two types of constraints, namely, hard constraints and soft constraints. Hard constraints refer to the constraints that must be met. For example, a user requests a group of VMS instead of a single VM. To ensure high availability, each VM needs to be placed in a separate fault domain (a server sharing a single point of failure) to avoid economic losses caused by the failure of the cloud computing center. Soft constraints refer to constraints that can or can not be satisfied, but satisfying constraints can significantly improve service quality. For example, in the process of neural network training, placing cloud virtual machines close to each other in the network structure can reduce network latency and speed up computing. During the processing of user requests, VMS that are geographically close to users can improve the communication speed between the cloud computing center and users and improve user experience.

This study mainly proposes a cloud resource allocation method based on deep reinforcement learning considering user needs. This study mainly considers assigning users to servers that are close to improve the service quality of users. However, if users are assigned to servers that are close to them, some servers may be congested and the waiting time will be long. Therefore, this study considers a cloud resource allocation method based on deep reinforcement learning. Deep reinforcement learning can determine the dynamic allocation of resources according to the current state of the system, thus maximizing the utilization efficiency of cloud resources and reducing user waiting time.

## 2.3 VM Migration Optimization and machine learning in Resource Management

Vm migration optimization in resource management is a complex problem, which involves many aspects such as resource allocation, load balancing, and power consumption optimization. Machine learning (ML) methods can help optimize the virtual machine migration decision process by predicting loads, resource requirements, and so on to develop migration strategies. Here, we can explore a simplified model that uses machine learning to optimize virtual machine migration.

(1) Problem definition

Let's say our goal is to minimize the energy consumption of the entire data center while ensuring that the performance of all virtual machines (VMs) is not affected. We can achieve this by intelligently migrating VMs to different physical machines (PMs).

(2) Feature selection

In order to train a machine learning model, we need to choose the right features. These characteristics may include:

Table 1. Table of optimized resource management indicators for VM migration

| Feature | Description |
| --- | --- |
| CPU usage of the VM | The percentage of CPU resources being used by the VM. |
| Memory usage of the VM | The amount of memory resources being used by the VM. |
| Storage I/O usage of the VM | The rate at which data is being read from and written to storage by the VM. |
| Network I/O usage of the VM | The rate at which data is being sent and received over the network by the VM. |
| Remaining CPU capacity of the PM | The amount of CPU resources that are not currently being used by any VM on the physical machine (PM). |
| Remaining memory capacity of the PM | The amount of memory resources that are not currently being used by any VM on the PM. |
| Energy consumption of PM | The amount of electrical power being consumed by the PM. |

Table 1 lists the key characteristics and metrics used for resource management and decision making during VM migration optimization. These characteristics and metrics are critical to understanding and evaluating virtual machine (VM) and physical machine (PM) performance, resource utilization, and power consumption. By analyzing this data, you can help optimize resource allocation in the data center, improve energy efficiency, and ensure that the performance of the service is not affected.

(3) Model selection

To solve this problem, we can consider using reinforcement learning (RL), especially Q-Learning or Deep reinforcement learning (DRL), such as Deep Q-Networks (DQN). These models can learn optimal migration strategies by interacting with the environment.

(4) Practical application

In a practical application, we need a more detailed environment model to simulate the dynamic interaction of VM and PM. In addition, the training of the model may require a large amount of data and computational resources. Depending on the specific needs, we may also need to consider using other ML models, such as decision trees, support vector machines, or neural networks, to predict the VM's resource requirements. Therefore, through continuous iteration and optimization, machine learning models can help us manage resources more efficiently and optimize the migration strategy of virtual machines to achieve the goal of reducing energy consumption and maintaining service performance.

## 3. VM RESOURCE OPTIMIZATION CASE AND METHODOLOGY

The essence of the online migration technology is that VMS can be migrated from one physical machine to another without stopping. Create VMS with the same configuration on the target physical machine, migrate all kinds of data, and quickly switch to a new VM on the target physical machine. During the migration process, user VMS can run normally for most of the time, and the last phase of the switchover process is very short, which does not interrupt user services and has little impact on user services running on VMS. Therefore, online migration is of great significance in realizing dynamic adjustment of cloud platform resources and troubleshooting.Therefore, this paper divides the online migration process into three stages: preparation stage, migration stage and switching stage.

Therefore, in the field of virtual machine migration optimization, the application of machine learning is gradually becoming a hot research spot, because it can help automate the optimization of resource allocation and migration decisions to improve the efficiency and performance of cloud computing environments. Here are some of the more innovative use cases that demonstrate the use of machine learning techniques in optimizing virtual machine migration:

### 3.1 Optimization of resource demand based on prediction

In this application case, the machine learning model is used to predict the future resource requirements (such as CPU, memory, etc.) of the virtual machine. By forecasting, the system can migrate VMS in advance to ensure effective resource allocation and avoid resource excess or insufficiency. This predictive approach can significantly reduce performance degradation due to resource reallocation and improve the overall efficiency of the cloud environment.

Typically, a time series prediction model, such as a long short-term memory network (LSTM) or gated cycle unit (GRU), can be used to predict a virtual machine's future resource requirements. The model diagram is as follows:

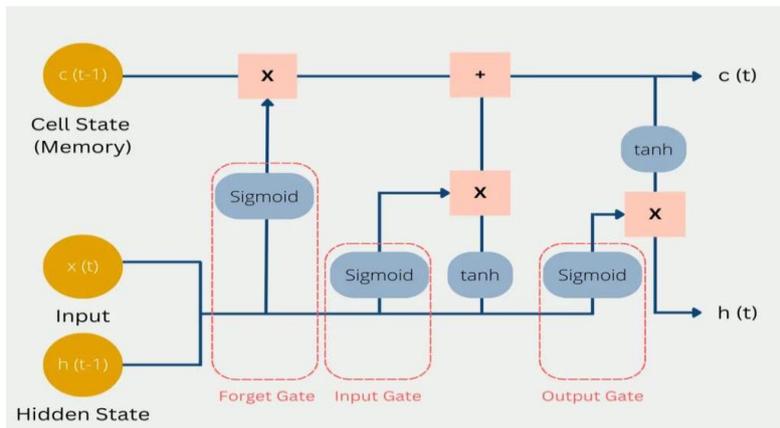

Figure 1. LSTM cloud computing resource optimization model

The LSTM (Long Short Term Memory) unit is designed to solve the problem of gradient disappearance or gradient explosion encountered by conventional recurrent neural networks (RNNS) when processing long sequence data. The core idea of LSTM is to maintain a long-term state or memory through special structural units. An LSTM unit consists of three main gates: a Forget Gate, an Input Gate, and an Output Gate, and a Cell State. These components work together to enable LSTM units to learn long-term dependencies in sequence data.

In the application of virtual machine resource demand forecasting, the output of the LSTM layer is fed into one or more fully connected layers (output layers), which are responsible for translating the high-dimensional output of the LSTM into specific predictions, such as CPU usage or memory requirements over a future period of time. The LSTM model is trained to learn patterns and dependencies in the sequence data, enabling the output layer to accurately predict future resource requirements.

Through this mechanism, LSTM can effectively capture long-term dependencies in time series data, providing a powerful predictive tool for optimizing virtual machine migration. This can not only improve the efficiency of resource allocation, but also help reduce performance issues caused by insufficient or excess resources.

Example function formula:

Let Xt be the input feature vector of time tt (for example, historical CPU and memory usage) and Yt be the predicted output of time tt (i.e. future resource requirements). LSTM units can be described by the following formula:

$$f_t=\sigma(W_f \cdot [h_{t-1},X_t]+b_f) f_t=\sigma(W_f \cdot [h_{t-1},X_t]+b_f) \quad (1)$$

$$i_t=\sigma(W_i \cdot [h_{t-1},X_t]+b_i) i_t=\sigma(W_i \cdot [h_{t-1},X_t]+b_i) \quad (2)$$

$$o_t=\sigma(W_o \cdot [h_{t-1},X_t]+b_o) o_t=\sigma(W_o \cdot [h_{t-1},X_t]+b_o) \quad (3)$$

$$\tilde{C}_t=\tanh(W_C \cdot [h_{t-1},X_t]+b_C) \tilde{C}_t=\tanh(W_C \cdot [h_{t-1},X_t]+b_C) \quad (4)$$

$$C_t=f_t*C_{t-1}+i_t*\tilde{C}_t C_t=f_t*C_{t-1}+i_t*\tilde{C}_t \quad (5)$$

$$h_t=o_t*\tanh(C_t) h_t=o_t*\tanh(C_t) \quad (6)$$

Where σ represents the sigmoid activation function, *represents the product of elements, W and b represent the model parameters, ht and Ct represent the hidden state and memory unit state of time tt respectively.

### 3.2 Dynamic migration decision optimization

Dynamic migration decision optimization involves using machine learning models to analyze the state of the cloud environment in real time and make virtual machine migration decisions to optimize resource utilization and reduce energy consumption. For example, deep reinforcement learning (DRL) can be used to learn the best migration strategy for different environmental states. This approach takes into account not only the current resource usage, but also the costs of the migration operation itself, such as migration time and performance impact, enabling more refined and efficient resource management.

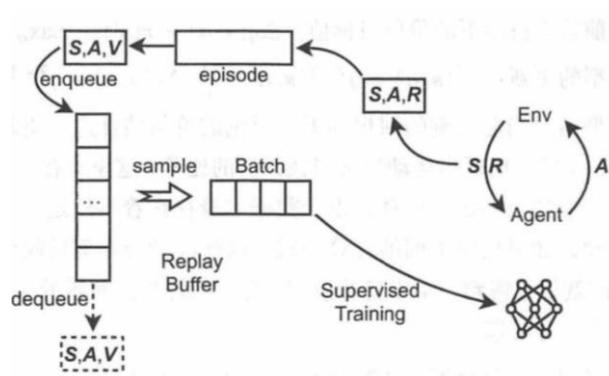

Figure 2.DQN model architecture decision optimization

The Deep Q Network (DQN) model is an algorithm that combines the principles of deep learning and reinforcement learning and is particularly suitable for solving decision optimization problems, such as resource management and optimization in cloud computing environments. DQN interacts with the environment and learns how to take action to maximize long-term rewards. In the context of virtual machine migration and resource management, DQN can be used to intelligently decide when and how to migrate virtual machines to optimize resource utilization and reduce energy consumption. The following are the key steps and principles for DQN to achieve resource management and optimization:

(1) Environmental modeling

In the DQN framework, the cloud computing environment is modeled as a Markov decision process (MDP), where each state represents the current configuration of the cloud environment, including the resource usage of each virtual machine, the load on the physical server, and so on. An action corresponds to a decision to migrate a virtual machine, such as from one physical server to another.

(2) Learning objectives

The goal of DQN is to learn a strategy, which is to choose rules of action in a given state in order to maximize future cumulative rewards. The reward function here needs to be defined in terms of resource management goals, such as improving resource utilization, reducing energy consumption, maintaining load balance, or reducing migration costs.

(3) Deep Q learning

DQN utilizes deep neural networks to approximate the Q function, the expected future reward for a given state and action. The input to the network is the state of the current environment and the output is the Q value of each possible action. Through interaction with the environment and feedback from reward signals, DQN constantly updates the network weights to more accurately estimate Q values.

(4) Experience playback

DQN uses experiential replay mechanism to break the temporal correlation between samples and improve the learning stability. It stores experienced state transitions (state, action, reward, next state) in the playback cache, and randomly draws small batches of past experiences to update the network, which helps the network learn from a wider range of experiences and avoid overfitting to recent or few experiences.

(5) Target network

In order to further stabilize the learning process, DQN introduces the concept of target network. The target network is a regularly updated copy of the Q network that is used to calculate the target Q value. This separate target network can reduce target movement during learning and make training more stable.

(6) Resource management and optimization decision-making

During the learning process, the DQN makes decisions based on the current state by selecting an action with the maximum Q value, such as choosing to migrate the virtual machine or keep the current configuration. As training progresses, the model becomes better at predicting which actions will lead to the best long-term outcomes, enabling efficient management and optimization of resources.

In this way, DQN can automatically learn how to make the best resource management and virtual machine migration decisions in complex cloud computing environments to improve the overall performance and efficiency of the system.

### 3.3 Energy consumption optimization

In cloud data centers, energy consumption is an important consideration. Using machine learning models to optimize virtual machine migration can significantly reduce energy consumption. By learning the usage patterns of virtual machines and the energy consumption characteristics of the data center, the machine learning algorithm can identify energy-saving migration strategies, such as migrating part of the load from a high-energy node to a low-energy node, or dynamically adjusting the energy use of the data center based on the load to achieve green computing.

**Example function formula:**

Let X be the input eigenvector and Y be the predicted energy consumption. MLP can be described by the following formula:

$$Y = f(W_n \cdot (f(W_{n-1} \ldots f(W_1 \cdot X + b_1) \cdots + b_{n-1})) + b_n \quad (7)$$

Where W and b represent the weight and bias of the network, f represents the activation function (such as ReLU or tanh), and n represents the number of layers of the network. By equipping these cases with relevant model diagrams or functional formulas, you can more intuitively show how machine learning can play a role in virtual machine migration optimization and how different technical approaches can be used to solve specific problems.

These use cases demonstrate the multiple potential of machine learning in optimizing virtual machine migration, from prediction and dynamic tuning to fault recovery and power management, machine learning technology is becoming a key force driving advances in cloud computing resource management. With the development and application of technology, more innovative and efficient solutions are expected in the future.

### 3.4 Deep reinforcement learning (DRL) and it in dynamic resource allocation

Deep reinforcement learning (DRL) combines the perceptual capabilities of deep learning with the decision making capabilities of reinforcement learning to provide a powerful approach to dynamic resource allocation and management. The DRL enables intelligent and automatic management of cloud computing resources and optimizes VM deployment, migration, and expansion to improve resource utilization and reduce power consumption. Below is an exploration of the application of DRL in dynamic resource allocation, as well as the challenges and potential solutions to using this technology.

(1) Application exploration of DRL in dynamic resource allocation

Environment modeling: First, the cloud computing environment needs to be modeled as a Markov decision process (MDP), where each state represents the current resource allocation and system state, each action represents a possible resource allocation decision, and the reward function is defined in terms of resource utilization efficiency and quality of service.

Strategy learning: DRL aims to learn a strategy of what actions to take in a given state to maximize long-term rewards. This includes learning how to dynamically adjust resource allocation based on current system load and resource usage, how to pre-allocate resources based on predicted changes in demand, and when to trigger the migration or scaling of virtual machines.

The ability to adapt and generalize: The ability of DRL models to learn from historical data and adapt to new, unseen environmental states is critical to dealing with the dynamics and uncertainties of cloud computing environments. The generalization ability of the model enables it to make effective decisions under multiple workloads and configuration changes.

Challenges and potential solutions for using DRL technology

(2) Challenge

Complexity of state space and action space: Cloud computing environments are often highly complex and the state space and action space can be very large, which makes the learning process difficult and computationally intensive.

Sparsity of reward signals: In practical applications, effective reward signals can be rare, making it difficult for models to quickly learn effective strategies.

The need for real-time decisions: Resource management decisions in cloud environments often need to be made in a short period of time, which challenges the speed of inference of DRL models.

(3) Solutions

State space and action space simplification: Simplify learning tasks by reducing the dimensions of state space and action space through feature engineering or using techniques such as autoencoders.

Reward Shaping: Use reward shaping techniques to increase the frequency of reward signals by introducing intermediate rewards to help models learn effective strategies faster.

Model lightweight and acceleration: Through the model compression, quantization and structure optimization technology to reduce the calculation requirements of the model, improve the speed of model inference, to meet the needs of real-time decision-making.

Transfer learning and meta-learning: Use transfer learning and meta-learning techniques to enable DRL models to leverage knowledge learned in other tasks or environments to accelerate the learning process in new environments.

Through these approaches, the application potential of DRL technology in dynamic resource allocation and management has been significantly enhanced, and despite the challenges, as the technology advances and more research is carried out, DRL will play an increasingly important role in intelligent cloud computing resource management.

## 4. CONCLUSION

This article takes an in-depth look at the application of machine learning and deep reinforcement learning (DRL) techniques in cloud resource management and virtual machine migration optimization, highlighting the importance of these advanced technologies in dealing with the dynamic changes and complexities of cloud computing environments. Through environment modeling, policy learning and adaptive capability enhancement, machine learning methods, especially DRL, provide effective solutions for dynamic resource allocation and intelligent migration of virtual machines. These technologies can help cloud service providers improve resource utilization, reduce power consumption, and improve service reliability and performance.

Although machine learning and DRL show great potential in cloud resource management, their practical application still faces many challenges, including complex state and action Spaces, sparsity of reward signals, and the need for real-time decision making. Effective strategies to address these challenges include simplifying state space and action space, rewarding shaping, model lightweight and acceleration, and accelerating the learning process with transfer learning and meta-learning techniques. These solutions help improve the learning efficiency and decision quality of DRL models, making them more suitable for the fast-paced and ever-changing cloud computing environment.

With the continuous progress of machine learning and deep reinforcement learning technology, combined with the rapid development of cloud computing technology, it is expected that the application of these technologies in cloud resource management and virtual machine migration optimization will be more extensive and in-depth. Researchers will continue to explore more efficient algorithms and models to further improve the accuracy and efficiency of decision making. In addition, with the integration of edge computing and the Internet of Things and other technologies, cloud computing resource management will face more new challenges and opportunities, and the scope and depth of application of machine learning and DRL technology will also expand, opening up new possibilities for building a more intelligent, efficient and reliable cloud computing service system.